\documentclass[prb,twocolumn,showpacs,amsmath,amssymb,superscriptaddress,floatfix]{revtex4}
\pdfoutput=1

\usepackage{ graphicx } 
\usepackage{ bm } 
\usepackage{ color }

\newcommand{\ua}{\uparrow}
\newcommand{\da}{\downarrow}

\newcommand{\be}{\begin{equation}}
\newcommand{\ee}{\end{equation}}
\newcommand{\bea}{\begin{eqnarray}}
\newcommand{\eea}{\end{eqnarray}}

\newcommand{\vk}{{\boldsymbol k}}

\begin{document}
\title{Charge and spin edge currents in $2d$ Floquet topological superconductors}
\author{ P. D. Sacramento }
\affiliation{ \textit CeFEMA,
Instituto Superior T\'ecnico, Universidade de Lisboa, Av. Rovisco Pais, 1049-001 Lisboa, Portugal }
\affiliation{ Centro Brasileiro de Pesquisas F\'{\i}sicas, Rua Dr. Xavier Sigaud 150, Urca 22290-180, 
Rio de Janeiro, RJ, Brazil}
\date{ \today }


\begin{abstract}
A time periodic driving on a topologically trivial system induces edge modes
and topological properties. In this work we consider triplet and singlet superconductors subject
to periodic variations of the chemical potential, spin-orbit coupling and magnetization,
in both topologically trivial and nontrivial phases, and study their influence on the
charge and spin currents that propagate along the edges of the two-dimensional system, for
moderate to large driving frequencies.
Currents associated with the edge modes are induced in the trivial phases and
enhanced in the topological phases.
In some cases there is a sign reversal of the currents as a consequence of the periodic driving.
The edge states associated with the finite quasi-energy states at the edge of the Floquet
zone are in general robust, while the stability of the zero quasi-energy states
depends on the parameters.
Also, the spin polarization of the Floquet spectrum quasi-energies is strong as for the
unperturbed topological phases. It is found that in some cases the unperturbed edge states
are immersed in a continuum of states due to the perturbation, particularly if the driving
frequency is not large enough. However, their contribution
to the edge currents and spin polarization is still significant.
\end{abstract}


\maketitle

\section{Introduction}

Topological systems have attracted great interest \cite{kane,zhang} and, 
specifically in the case of topological superconductors \cite{alicea}, 
a great effort has been put towards prediction and
detection of Majorana fermions \cite{mourik}. The robustness of topological
systems to perturbations
is a key feature with wide potential applications, 
such as the edge state Majorana fermions as building blocks
for topological quantum computation, due to their non-abelian statistics.
It has been shown that topological systems are quite robust to a 
quantum quench, as exemplified by the toric code model \cite{tsomokos,hamma}. 
It has also been shown recently that, in an infinite system, 
the Chern number can not be changed by a unitary
evolution, even though it is possible to change the Bott index and topology 
if the system has a finite size \cite{alessio}, as obtained earlier
in the case of quenches in topological superconductors \cite{pre}.
Therefore, the response to a time dependent perturbation of the edge states in finite systems 
may not be protected by topology.
Furthermore, quenches in superconducting systems with topological properties, 
performed self-consistently \cite{dzero3}, showed the importance of the topological
properties in the evolution of the system \cite{sen1} and raised questions regarding the
survival of the topological order to the quench \cite{tsomokos,scheurer,dzero4,dzero5}. 
So the issue is not resolved and is attracting considerable attention.

Several examples that host Majorana fermions as edge states are provided by several
superconducting p-wave systems 
\cite{kitaev,pip} and various
other systems that mix superconducting order (eventually by proximity effects)
with Zeeman fields and/or spin-orbit coupling \cite{sato} or tunneling of Cooper pairs \cite{fukane}. 
A recent proposal for the existence of localized Majorana edge states at the ends of a magnetic
chain in contact with a conventional superconductor \cite{bernevig1} has been confirmed
experimentally \cite{science} (there is however some debate because of their unexpected
very localized nature \cite{glazman,dassarma}) and it has been shown, combining topological with non-topological
regions in wires, that it also provides a good basis of
states to implement the braiding properties of the Majorana fermions \cite{necklace}, with
potential interest in quantum computation. The fermion exchange in a unidimensional system
recently proposed \cite{dwallfranz} is also possible to implement with a magnetic chain.
Various other proposals have been presented for the existence of Majorana zero energy states
including recently multiband systems \cite{viola1,viola2,mucio}, where the non-trivial interband
coupling leads to interesting topological properties, as shown also in the context of
topological Kondo insulators \cite{dzero1,dzero2}. 

In higher dimensions the edge modes are in general propagating in the directionss parallel
to the edge and localized in the perpendicular direction.
Their nature as protected states suggests the existence of spontaneous currents, either charge or
spin currents, that have aroused considerable interest in several contexts such as
in the context of spintronics, as a means to generate and manipulate spin currents.
The anomalous Hall effect due to the presence of spin-orbit coupling
has been considered \cite{sigrist1,ahe,epl} as well as the appearance of currents
due to magnetic impurities \cite{ahe,abs}.
The study of the currents
generated at interfaces between topological insulators and a supercondutor has been
considered \cite{sen3,sen4} such as the charge and spin conductances. 
A manipulation using electrical means has been proposed \cite{andrei} as well as their
spin correlations using a gate potential \cite{tanaka1}. 
Chiral $p-$wave superconductors are expected to show spontaneous edge currents.
Some other chiral systems were, however,
shown to not have spontaneous currents \cite{sigrist2,kallin2,kallin3} 
but, with some admixtures of different
pairings, these arise together with some non-trivial spin polarization \cite{timm}.
Even though the topic has been studied for some time,
the experimental verification of the spontaneous generation of edge currents is still
a topic of considerable attention. It is expected that a strong candidate in the class of
chiral p-wave superconductors $Sr_2RuO_4$
will exhibit spontaneous currents \cite{maeno1,kallin1}. 
Even though the material $Sr_2RuO_4$ has been shown to break time reversal symmetry
due to the existence of the Kerr effect \cite{kerr}, no edge surrents have been found
\cite{curran,kirtley} even though evidence for the edge states has been seen by 
in-plane tunneling spectroscopy \cite{maeno2}. 
Various proposals have been presented
to explain the absence of the edge currents \cite{simon} using the result of an effective
high Chern number \cite{oshikawa,volovik1} or effects of disorder \cite{kallin2}. 
Also, it has been proposed that the absence of the edge currents may be due to a 
spin singlet pairing instead of the generally assumed p-wave triplet pairing \cite{singlet}.

As mentioned above, it is interesting to study the
robustness of the edge states to time dependent perturbations.
In the context 
of the Creutz ladder, it was shown that the presence of edge states modifies the
process of defect production expected from the Kibble-Zurek mechanism, leading
in this problem to a scaling with the change rate with a non-universal critical
exponent \cite{bermudez1}. A similar result was obtained for the one-dimensional
superconducting Kitaev model, where it was shown that, although bulk states follow
the Kibble-Zurek scaling, the produced defects for an edge state quench are quite
anomalous and independent of the quench rate \cite{bermudez2,pre}. 
The behavior of edge states under an abrupt quantum quench has also been considered
very recently in the context of a two-dimensional topological insulator \cite{bhz},
where it was found that, in the sudden transition from the topological insulator to
the trivial insulator phase, there is a collapse and revival of the edge states \cite{patel}.
Similar results were obtained for the one-dimensional Kitaev model \cite{rajak}, and also
studying the signature of the Majoranas in the entanglement spectrum \cite{chung}.
Their dynamical formation and manipulation has been considered in \cite{perfetto} and
\cite{scheurer}.
The effect of a sudden quench of the parameters of the Hamiltonian 
of a two-dimensional triplet superconductor
has also been studied, and the robustness of the edge states was considered \cite{pre}.
In general it was found that the edge states decay due to the quench, even though
in some cases they are quite robust, such as in the case of weak spin-orbit coupling,
or when there are matching momentum states in the initial and final states.
The effect of parity blocking on the dynamics of the edge modes has been considered recently
in which case the dynamics is restricted if there is a change in fermion parity accross
the quench \cite{blocking}.
On the other hand, the Majorana zero modes lead to some universal non-equilibrium signature
in the Loschmidt echo with an universal exponent associated with the algebraic decay
\cite{moore1}. Also, the dynamics of the tunneling into non-equilibrium edge states has been
proposed as a possible signature of the existence of these states \cite{moore2}.
Non-equilibrium situations also may allow the transport of Majorana edges states using extended
gapless regions with a small but finite overlap with the Majoranas \cite{dutta1}. Their effect
has also been considered in \cite{sen2}.

While quenches, either abrupt or slow, in general destabilize the edge states, topological phases 
can be induced by periodically driving the Hamiltonian
of a non-topological system, such as shown before in topological insulators \cite{oka,floquet0,lindner} 
and in topological superconductors, with the appearance of Majorana fermions \cite{jiang,luo,xiaosen,viola}.
Their appearance in a one-dimensional p-wave superconductor was studied in Ref. \cite{liu}
and in Ref. \cite{manisha} introducing external periodic perturbations; the case of
intrinsic periodic modulation was also considered \cite{foster}.
The periodic driving leads to new topological states \cite{lindner}, and to a
generalization of the bulk-edge correspondence, that reveals a richer structure \cite{prx,balseiro} as
compared with the equilibrium situation \cite{symmetry,delplace}. 
Similarly, in topological superconductors
new phases may be induced and manipulated due to the presence of the 
periodic driving \cite{platero,hexagonal,liu}. 

In general the problem is complex due to problem of dissipation through coupling to
baths. Their effect have
recently been considered and their detection has been proposed
using transport properties \cite{mitra1,mitra2,hanggi},
as well as magnetization
signatures using the magnetic fields the currents should produce \cite{moore3}.

In this work we compare the charge and spin edge currents of the unperturbed
Hamiltonian with those of the perturbed Hamiltonian for triplet superconductors.
We consider a system with no coupling to its environment except for the coupling
to the periodic perturbation. To simplify, we consider perturbations where
only one of the parameters of the Hamiltonian is periodicaly changed, such
as the hopping, spin-orbit coupling, chemical potential or magnetization.
We mainly focus on the last two parameter changes since are easier to implement
experimentally.
In section II we review the Hamiltonian considered and in section III we review the
theory of the Floquet states that result from a periodic perturbation.  In
section IV we illustrate the method by presenting solutions of the quasi-energies
for different cases, as a function of the perturbation amplitude and frequency, and
verify the convergence of the truncation procedure of the Hamiltonian matrix of the
Floquet problem. In section V we present results for the edge currents and the spin
polarization of the states of system. We conclude with section VI.

\section{Unperturbed topological superconductor}

We consider two-dimensional superconductors that display topological phases.
Standard examples are triplet superconductors
with $p$-wave symmetry. We also consider the presence of spin-orbit coupling and a Zeeman term
(magnetization that may be due to proximity effect or intrinsic).
In the case of a non-centrosymmetric system, since parity is no longer conserved,
a spin singlet pairing is also possible. Therefore we also consider an admixture of
a conventional $s$-wave component. Even without a triplet pairing component a conventional
superconductor may also display topological properties in the presence of magnetization.

The model considered here was studied in various refserences before such as in Refs. \cite{sato,epl}.
We write the Hamiltonian for the bulk system as
\begin{eqnarray}
\hat H = \frac 1 2\sum_\vk  \left( {\boldsymbol \psi}_{\vk}^\dagger ,{\boldsymbol \psi}_{-\vk}   \right)
\left(\begin{array}{cc}
\hat H_0(\vk) & \hat \Delta(\vk) \\
\hat \Delta^{\dagger}(\vk) & -\hat H_0^T(-\vk) \end{array}\right)
\left( \begin{array}{c}
 {\boldsymbol \psi}_{\vk} \\  {\boldsymbol \psi}_{-\vk}^\dagger  \end{array}
\right)
\label{bdg1}
\end{eqnarray}
where $\left( {\boldsymbol \psi}_{\vk}^{\dagger}, {\boldsymbol \psi}_{-\vk} \right) =
\left( \psi_{\vk\ua}^{\dagger}, \psi_{\vk\da}^\dagger ,\psi_{-\vk\ua}, \psi_{-\vk\da}   \right)$
and
\begin{equation}
\hat H_0=\epsilon_\vk\sigma_0 -M_z\sigma_z + \hat H_R\,.
\end{equation}
Here, $\epsilon_{\boldsymbol{k}}=-2 \tilde{t} (\cos k_x + \cos k_y )-\mu$
is the kinetic part, $\tilde{t}$ denotes the hopping parameter set in
the following as the energy scale ($\tilde{t}=1$),
$\mu$ is the chemical potential,
$\boldsymbol{k}$ is a wave vector in the $xy$ plane, and we have taken
the lattice constant to be unity. Furthermore, $M_z$
is the Zeeman splitting term responsible for the magnetization,
in $\tilde{t}$ units.
The Rashba spin-orbit term is written as
\begin{equation}
\hat H_R = \boldsymbol{s} \cdot \boldsymbol{\sigma} = \alpha
\left( \sin k_y \sigma_x - \sin k_x \sigma_y \right)\,,
\end{equation}
 where $\alpha$ is measured in the same units.
The matrices $\sigma_x,\sigma_y,\sigma_z$ are
the Pauli matrices acting on the spin sector, and $\sigma_0$ is the
$2\times 2$ identity.
The pairing matrix reads \cite{Sigrist}
\begin{equation}
\hat \Delta = i\left( {\boldsymbol d}\cdot {\boldsymbol\sigma} \right) \sigma_y =
 \left(\begin{array}{cc}
-d_x+i d_y & d_z+\Delta_s \\
d_z-\Delta_s & d_x +i d_y
\end{array}\right)\,.
\end{equation}
where $\Delta_s$ is the symmetric part and the vector $\boldsymbol d$ parametrizes the
anti-symmetric part.

The energy eigenvalues and eigenfunction may be obtained solving the Bogoliubov-de Gennes equations
\be
\label{bdg2}
\left(\begin{array}{cc}
\hat H_0(\vk) & \hat \Delta(\vk) \\
\hat \Delta^{\dagger}(\vk) & -\hat H_0^T(-\vk) \end{array}\right)
\left(\begin{array}{c}
u_n\\
v_n
\end{array}\right)
= \epsilon_{\boldsymbol{k},n}
\left(\begin{array}{c}
u_n\\
v_n
\end{array}\right).
\ee
The 4-component spinor can be written as
\be
\left(\begin{array}{c}
u_n\\
v_n
\end{array}\right)=
\left(\begin{array}{c}
u_n(\boldsymbol{k},\uparrow) \\
u_n(\boldsymbol{k},\downarrow) \\
v_n(-\boldsymbol{k},\uparrow) \\
v_n(-\boldsymbol{k},\downarrow) \\
\end{array}\right) .\ee

We consider either a superconductor with
$\boldsymbol{d}=d (\sin k_y,-\sin k_x,0)$
or
$\boldsymbol{d}=d_z (0,0,\sin k_x-i \sin k_y)$
with or without a contribution from the local s-wave pairing $\Delta_s$.
The first case applies if the spin-orbit is strong. In this case the pairing 
is aligned \cite{Sigrist2} along the spin-orbit vector $\boldsymbol{s}$.
This case is denoted by strong coupling case.
Relaxing this restriction allows that the two vectors are not aligned. This case
is denoted by weak spin-orbit coupling and has been considered before in the
context of the anomalous Hall effect and the calculation of the Hall conductance \cite{ahe,epl}.
In the strong-coupling case $\boldsymbol{d}=(d_x,d_y,d_z) = ( d / \alpha ) \boldsymbol{s} $.
The second case, usually called $p+ip$ superconductor, 
has been the focus of great attention since it has spontaneous chiral
states due to its time-reversal breaking structure. In contrast, in the absence of the Zeeman
term the first case does not break time-reversal symmetry (TRS) and in the topological phases
the edge states have a helicoidal structure.
The system then belongs to the symmetry class DIII where the topological invariant is
a $\mathbb{Z}_2$ index \cite{symmetry}.
If the Zeeman term is finite or in the case of the $p+ip$ pairing, TRS is broken and the system belongs
to the symmetry class D.
The topological invariant that characterizes this phase is the first Chern number $C$,
and the system is said to be a $\mathbb{Z}$~topological superconductor.

\begin{figure}
\includegraphics[width=0.9\columnwidth]{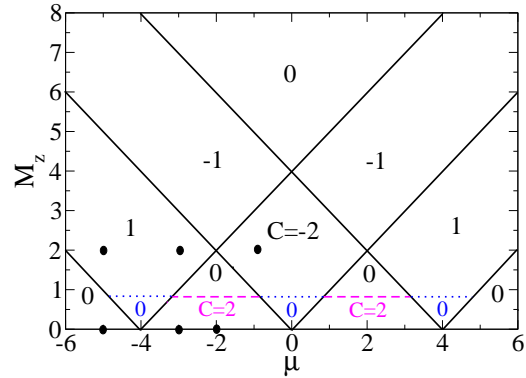}
\caption{\label{fig1}
(Color online) Phase diagram for the spin triplet superconductor
with ${\boldsymbol d}=(d_x,d_y,0)$. The phase diagramfor a $s$-wave
superconductor is very similar except in the vicinity of zero magnetization
(limited approximately by the dotted lines)
where the system is topologically trivial. In the case of a triplet
superconductor with ${\boldsymbol d}=(0,0,d_z)$ with a complex $z$-component,
the phase diagram is also similar with the exception of a region around zero
magnetic field 
(limited by dashed lines)
that is topological, with Chern number 2.}
\end{figure}

The Chern number is
obtainable as an integral of the Berry curvature over the Brillouin zone \cite{xiao,Fukui}.
Summing over the occupied bands the Chern number has been calculated \cite{sato,epl}.
It is convenient to calculate the Chern number by computing the
flux of the Berry curvature over plaquetes in the Brillouin zone \cite{Fukui}.

The results in the parameter space for the various pairing symmetries are shown in Fig. \ref{fig1}
using the parameters $\alpha=0.6$, $d=0.6$ and in the case of strong spin-orbit coupling. 
In this case, changing
these parameters only leads to a quantitative change of the shape of the energy bands 
and no qualitative changes
are observed in the topological properties. As shown before \cite{sato}, the spin-orbit coupling
does not change the topology.
The phase diagram for the case of a $s$-wave superconductor is very similar except for values of
the magnetization smaller than the amplitude of $\Delta_s$. In these regimes the Chern number vanishes
and the system is topologically trivial for all values of the chemical potential.
Something similar happens in the case of the $p+ip$ superconductor in the sense that some regimes
at values of the magnetization smaller than the pairing amplitude, $d_z$, the Chern number is
different from the case of the $d_x,d_y$ pairing. In these regimes it is finite, $C=2$, due to
the broken TRS, even in the absence of a finite magnetization.
However, for some values of the chemical potential,  a finite magnetization will change the
Chern number to zero (even though there are unprotected edge states).

Due to the bulk-edge correspondence if the system is placed in a strip geometry and the system
is in a topologically non-trivial phase, there are robust edge states, in a number of pairs given by
the Chern number, if time reversal symmetry is broken. There are also counterpropagating edge states
in the $Z_2$ phases even though the Chern number vanishes, as in the spin Hall effect. In these
phases time reversal symmetry is preserved and the Kramers pairs of edge states give opposite
contributions to the Chern number. Interestingly, turning on the magnetization (Zeeman field)
time reversal symmetry is broken and the edge states are no longer topologically protected. However,
it was found that, even in regimes where $C=0$, there are edge states, reminiscent of the edge
states of the $Z_2$ phases, as long as the chemical potential satisfies $|\mu|<4\tilde{t}$.
The regimes where the magnetization is zero and $|\mu|>4\tilde{t}$ are topologically trivial and
in a strip geometry have no edge states.

\section{Periodic driving}

The time evolution of a state under the influence of a time dependent
Hamiltonian is given by
\be
i  \frac{\partial}{\partial t} \psi(k,t)=H(k,t) \psi(k,t)
\ee
where $k$ is the momentum, $t$ the time and $\hbar=1$.
We can decompose the Hamiltonian in two terms a time independent one,
$H(k)$,
and an extra term due to the external time-dependent perturbation, that
we want to take as periodic with a given frequency, $\omega$,
\be
H(k,t)=H(k)+f(\omega t) H_d(k)
\ee
Here $H_d(k)$ is of the form of the unperturbed Hamiltonian but with
only one non-vanishing term.
Looking for a solution of the type 
\be
\psi(k,t)=e^{-i \epsilon(k)t} \Phi(k,t)
\ee
and using that $\Phi(k,t)=\Phi(k,t+T)$, where $T$ is the period ($\omega=2\pi/T$)
and $f(\omega t)=f(\omega(t+T))$, one gets that
\be
\left( H(k,t)-i\frac{\partial}{\partial t} \right) \Phi(k,t)=\epsilon(k) \Phi(k,t)
\ee
The time-independent quasi-energies $\epsilon(k)$ are the eigenvalues of
the operator $H(k,t)-i\frac{\partial}{\partial t}$ and the function $\Phi(k,t)$ the
eigenfunction.
Since this function is periodic, we can expand it as
\be
\Phi(k,t)=\sum_m \phi_m(k) e^{i m \omega t}
\ee
Inserting this expansion in equation we obtain the eigensystem
\be
\sum_{m^{\prime}} H_{m m^{\prime}}(k) \phi_{m^{\prime}} (k) = \epsilon(k) \phi_m (k)
\ee
The Hamiltonian matrix is given by
\be
H_{m m^{\prime}}(k) = \delta_{m m^{\prime}} m \omega +
\frac{1}{T} \int_0^T dt e^{-i m \omega t} H(k,t) e^{i m^{\prime} \omega t}
\ee

Choosing a perturbation of the type $f(\omega t)=\cos (\omega t)$ the second term
of the Hamiltonian matrix reduces to $1/2 \left( \delta_{m^{\prime}+1,m} + \delta_{m^{\prime}-1,m} \right)$.

The time evolution of the state is then obtained solving for the quasi-energies, $\epsilon(k)$,
and the functions $\phi_m(k)$ diagonalizing the infinite matrix
\begin{widetext}
\be
\label{hmm}
\left(\begin{array}{ccccccc}
\cdots & \cdots & \cdots & \cdots & \cdots & \cdots & \cdots \\
\cdots & (m-2)\omega +H(k) & \frac{1}{2} H_d(k) & 0 & 0 & 0 & \cdots \\
\cdots & \frac{1}{2} H_d(k) & (m-1)\omega + H(k) & \frac{1}{2} H_d(k)& 0 & 0 &\cdots \\
\cdots & 0 & \frac{1}{2} H_d(k) & m\omega +H(k) & \frac{1}{2} H_d(k) & 0 & \cdots \\
\cdots &0 & 0 &   \frac{1}{2} H_d(k) & (m+1)\omega +H(k) & \frac{1}{2} H_d(k) & \cdots \\
\cdots & 0 & 0 & 0  & \frac{1}{2} H_d(k) & (m+2)\omega + H(k) & \cdots \\
\cdots & \cdots & \cdots & \cdots & \cdots & \cdots & \cdots 
\end{array}\right)
\ee
\end{widetext}
The matrix can be reduced if the frequency is high enough and only a few values of
$m$ are needed. 
The photon-dressing of the band structures due to the mixing of the
bands is the important effect we consider \cite{light}.
We will neglect here any photon emission/absorption processes that
affect the occupation numbers of the electrons. Equivalently one may consider that
the photons are off-resonance. 
The effects of creation/absorption of photons if the driving frequency
is not large enough were discussed before \cite{mitra2}.
The matrix $H_d(k)$ will be chosen as the unperturbed Hamiltonian,
where only one of the parameters, hopping, chemical potential, spin-orbit coupling
or magnetization will be considered to vary with time. The first three parameters
preserve time reversal symmetry while the magnetization naturally breaks time
reversal symmetry if the unperturbed Hamiltonian is in a regime with vanishing
magnetization. Emphasis will be placed on the effects of varying the chemical potential
or the magnetization which are easilly tuned externally. In this last case it has
been determined before \cite{manisha} that even though the low energy states have a very low energy,
they may not be strictly Majorana fermions since the eigenvalues of the Floquet
operator (time evolution operator over one time period) are not strictly $\pm 1$.

Due to the periodicity of the eigenfunctions, $\Phi(k,t+T)=\Phi(k,t)$, the action of the evolution
operator, ${\cal U}(t)$, on a state over a period, $T$, leads to the same state minus a phase 
\be
|\psi(T)\rangle = {\cal U}(T) |\psi(0) \rangle = e^{-i \epsilon T} |\psi(0) \rangle
\ee
Therefore, the quasi-energies are defined minus a shift of a multiple of $w=2\pi/T$, and
we can restrict the quasi-energies to the first Floquet zone, defined by the interval
$-w/2\leq \epsilon \leq w/2$. States with quasi-energies $\epsilon=w/2$ and $\epsilon=-w/2$
are therefore equivalent and there is a reflection of any bands as one exits the Floquet
zone from above (or below) and as one enters from below (or above). Considering the particle-hole
symmetry of a superconductor, $\gamma_{-\epsilon}=\gamma_{\epsilon}^{\dagger}$ and the equivalence
between the energies $\epsilon=-w/2,w/2$ one expects a new type of Majorana mode in addition
to any zero energy states, the usual Majorana modes.

For the large enough driving frequencies considered here, the frequency description considered
is particularly convenient. In the regime of small driving frequencies a time description is
more convenient.

\section{Energy bands and quasi-energies}

\begin{figure*}
\includegraphics[width=0.7\textwidth]{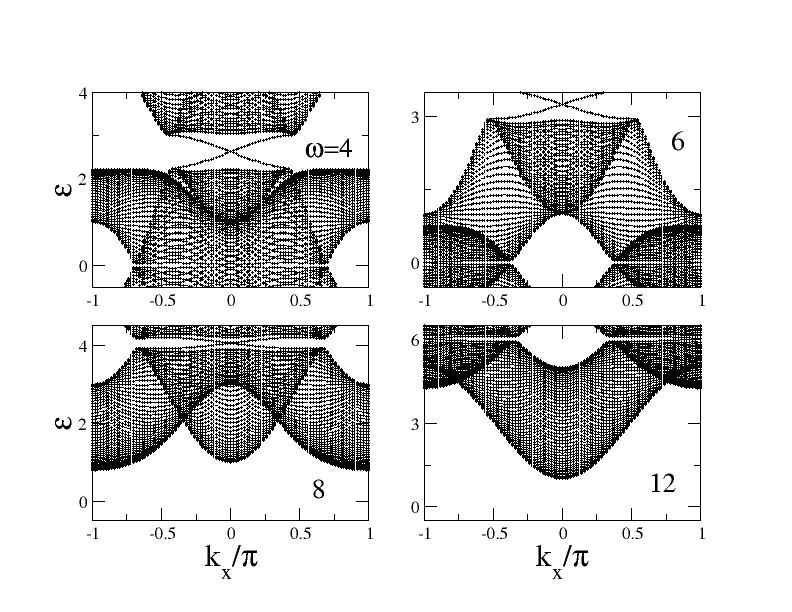}
\caption{\label{fig2}
Floquet spectra for $l=1$ for a $d_x,d_y$ triplet superconductor
in a topologically trivial phase with
$\alpha=0,M_z=0,\mu=-5,\Delta_s=0,d_z=0,d=0.6$
where the hopping is changed
with time with different frequencies $w=4,6,8,12$. The periodic driving is $t_d \cos wt$ with 
$t_d=2$. Note that the spectrum is symmetric around zero quasi-energy.} 
\end{figure*}

\begin{figure*}
\includegraphics[width=0.7\textwidth]{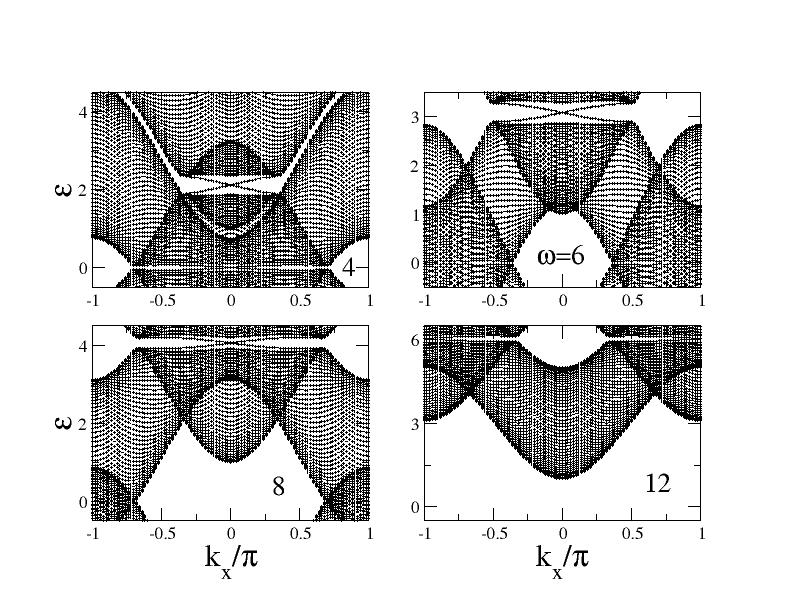}
\caption{\label{fig3}
Floquet spectra for $l=1$ for a $d_x,d_y$ triplet superconductor
in a topologically trivial phase with
$\alpha=0,M_z=0,\mu=-5,\Delta_s=0,d_z=0,d=0.6$
 where the magnetization is changed
with time with different frequencies $w=4,6,8,12$. The periodic driving is $M_{zd} \cos wt$ with 
$M_{zd}=2$.} 
\end{figure*}

\begin{figure*}
\includegraphics[width=0.7\textwidth]{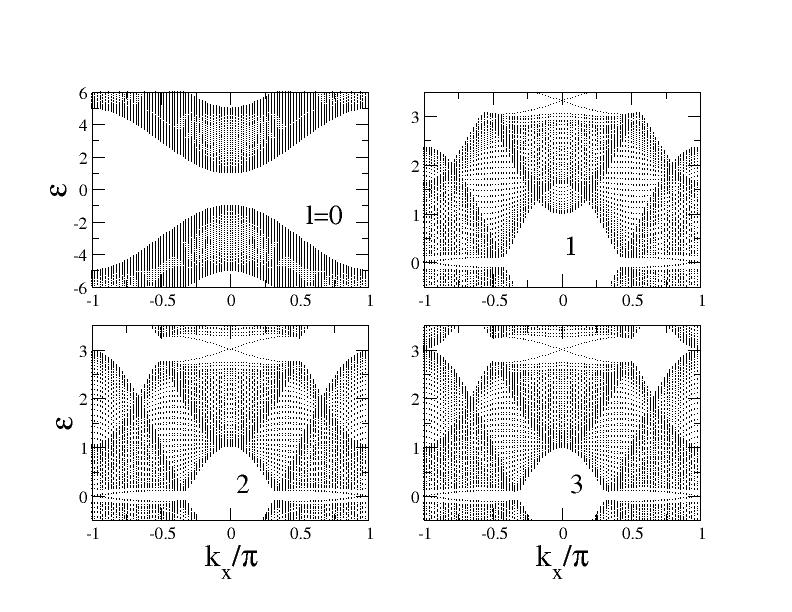}
\caption{\label{fig4}
Evolution of Floquet spectra for various values of $l$, $l=0,1,2,3$ for a $d_x,d_y$ triplet superconductor
in a topologically trivial phase
$\alpha=0.1,M_z=0,\mu=-5,\Delta_s=0.1,d_z=0,d=0.6$
 where the magnetization is changed
with time with frequency $w=6$. The periodic driving is $M_{zd} \cos wt$ with 
$M_{zd}=4$. The case $l=0$ is the unperturbed Hamiltonian. The results show the stability of two sets
of Majorana modes both at zero energy and at the limit of the Floquet zone.} 
\end{figure*}

\begin{figure*}
\includegraphics[width=0.7\textwidth]{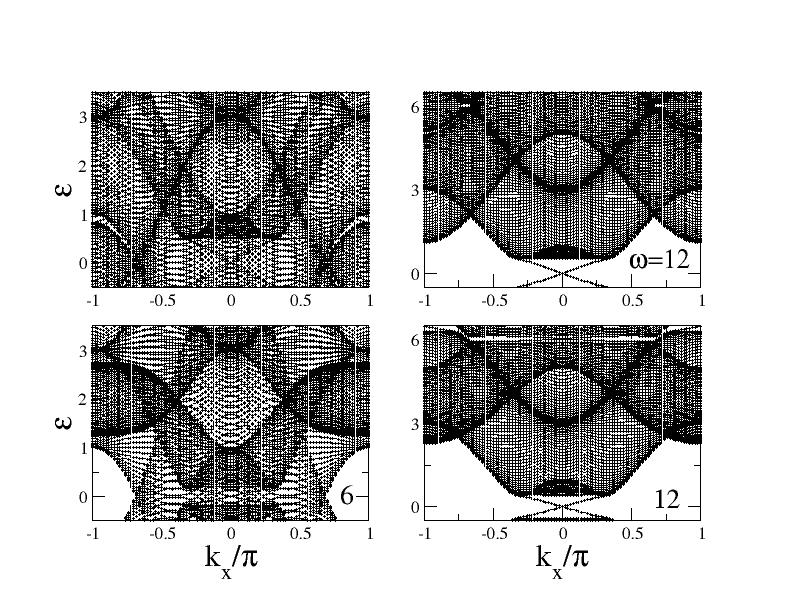}
\caption{\label{fig5}
Evolution of Floquet spectra for $l=1$ for a $d_x,d_y$ triplet superconductor
in a topologically non-trivial phase
$\alpha=0,M_z=2,\mu=-5,\Delta_s=0,d_z=0,d=0.6$
where the hopping is changed
with time with frequencies $w=6,12$. The periodic driving is $\tilde{t}_{d} \cos wt$ with 
$\tilde{t}_{d}=0.5$ (top row) and $\tilde{t}_d=2$ (bottom row). 
} 
\end{figure*}

The energy spectra of the unperturbed superconductors both in the trivial and
in the topological phases have been determined before. The solutions for the
quasi-energies of the perturbed Hamiltonians lead to bands that have a similar
structure. Here we focus on the energy and quasi-energy spectra in a ribbon geometry to give
evidence to the states along the edges of the system. 

At large frequencies, $w>4\tilde{t}$, the size of the truncated matrix is relatively
small and the quasi-energies and physical properties (calculated over the
first Floquet zone) converge fast for small values of $m$.
Considering $m=0$ one reproduces the Hamiltonian of the unperturbed superconductor.
The first approximation for the driven system is obtained considering $m=1,0,-1$, then $m=2,1,0,-1,-2$ and so on.
One may therefore use a short notation for the number of terms considered in the diagonalization
of the Hamiltonian matrix by using $l=0,1,2,3,\cdots$. The unperturbed case is denoted by $l=0$
and the perturbed cases by $l=1,2,\cdots$ (considering that we are using $2l+1$ states). 
If the frequency $w$ is small, one needs to consider
large values of $l$ and the problem of finding the edge states in a ribbon geometry quickly
becomes heavy computationally. Increasing the value of the frequency it is easy to find that
it is enough to consider $l=2$, since taking $l=3$ leads to very similar results, with a good
accuracy. 

To illustrate the structure of the quasi-energies we consider first the case of $l=1$ and a
$d_x,d_y$ superconductor in a topologically trivial phase. 
The results are shown in Figs. \ref{fig2} and \ref{fig3} where we
consider periodic drivings in the hopping and in the magnetization for moderate
couplings of $\tilde{t}_d=M_{zd}=2$. 
We consider frequencies $w=4,6,8,12$.
As the frequency increases 
edge states
appear at the border of the Floquet zone around $w/2$ (and $-w/2$). As the frequency increases
further there is a continuum of states that fills the gap around zero energy.
On the other hand, at higher energies close to the Floquet zone edge, as
the frequency increases the edge states appear on top of the continuum and prevail as
the gap around $w/2$ opens.
Therefore, as a consequence of the periodic driving edge states appear due to the Floquet
spectrum in a system that is trivial if unperturbed. 
The edge states around $w/2$ visible for the higher frequencies are robust and 
converge as $l$ increases. The two type of perturbations couple to 
states differently particularly in the spin structure. Also, the hopping preserves
time reversal symmetry while the magnetization breaks time reversal symmetry.
However, their effects at this order are very similar with respect to the edge states.
The main difference is the response of the bulk states to the two types of
perturbations, particularly seen at moderate frequencies where in the case of
the magnetization perturbation the continuum appears at the
energies where the edge states appear, even though at different momenta.

In Fig. \ref{fig4} we show the convergence of the truncation procedure of
the Hamiltonian matrix for a high value of the frequency $w=6$. We show results
for $l=1,2,3$ and see that the results for $l=2$ and $l=3$ are very similar. 
For a smaller frequency value, $w=4$, the convergence is also fast.
Moreover,
for the set of parameters considered, with a relatively high coupling of $M_{zd}=4$ we see
that, in addition to a fast convergence and effectiveness of the truncation procedure,
besides the edge Majorana states close to the Floquet boundary robust Majoranas
also appear at zero energy. For smaller frequencies and weak couplings the convergence
is in general slower. As a consequence, and to reduce the computational effort, we consider
a frequency $w=6$ in most results and limit the results to $l=2$.

It is also interesting to determine the effect of the periodic driving when the system
is originally in a nontrivial topological phase. 
In Fig. \ref{fig5} we show the effect of driving the hopping with different frequencies
$w=6,12$. The results are shown for $l=1$. 
We find that the continuum
fills completely the gap, as illustrated in the case of $w=6$. Also, the same happens at the
edge of the Floquet zone, particularly if the coupling is small $\tilde{t}_d=0.5$.
Increasing the coupling to $\tilde{t}_d=2$ an edge state emerges at the boundary of the
Floquet zone. Also both for moderate ($w=6$) and high frequencies ($w=12$) edge states
appear near zero energy.   
However, the influence of the bulk states is considerable and at moderate frequency $w=6$
they are superimposed on the edge states.

The topology of the system in the presence of the periodic driving may also be
considered in terms of the system in a torus. As mentioned above, the classification
is richer and the bulk-edge correspondance is more subtle. It has been proposed \cite{prx}
that one may calculate the Chern number of the occupied quasi-energy bands, as for the
unperturbed system. This works if there are clear gaps between the quasi-energy bands.
In the context of this work, this will work in general at high frquencies. In most other
cases, even though the analysis in the stripe geometry reveals edge states, these are
often mixed with continuum states that close the gaps between bands, preventing the
calculation of the Chern number of these bands. Also, due to the periodicity in Floquet
quasi-energy space, we may have bands with Chern number that vanishes if the number of edge
modes that enter the band equals the number of modes that leave it \cite{prx}.
Since our aim here are the currents associated with the edge modes, it is more enlightening
to look at the band structure in the stripe geometry, since both the edge and the bulk states
(projected along one spatial direction) are found.

\section{Currents}

We consider a finite system of dimensions $N_x \times N_y$.
We apply periodic boundary conditions along the $x$ direction and use a momentum representation
and open boundary conditions along the transverse direction, $y$, solving the problem
in a ribbon of width $N_y$.
Writing
\be
\psi_{k_x,j_y,\sigma} =  \frac{1}{\sqrt{N_x}} \sum_{j_x} e^{-i k_x j_x}
 \psi_{j_x,j_y,\sigma}\,,
\nonumber
\ee
we may rewrite the Hamiltonian matrix in terms of
these operators.
The diagonalization of this Hamiltonian involves the solution of a $(4 N_y) \times (4 N_y)$ eigenvalue problem
for each momentum $k_x$.
The energy states include states in the bulk and states along the edges. The eigenstates give us directly the
wave functions in real space.

In this representation the states are column vectors that with no external time dependent perturbation is of the type
\begin{widetext}
\be 
\left(\begin{array}{ccccccccc}
u(k_x,1,\uparrow) &
u(k_x,1,\downarrow) &
v(-k_x,1,\uparrow) &
v(-k_x,1,\downarrow) &
\cdots &
u(k_x,N_y,\uparrow) &
u(k_x,N_y,\downarrow) &
v(-k_x,N_y,\uparrow) &
v(-k_x,N_y,\downarrow) 
\end{array} \right)^T
\ee
\end{widetext}
The time evolved states due to the perturbation are the result of the diagonalization of the truncated
Floquet matrix and each component (${\cal U}=u$ or ${\cal U}=v$) is of the form
\be
{\cal U}(k_x,j_y,\sigma,t) = \sum_m e^{i m \omega t} \phi_m(k_x,j_y,\sigma)
\ee

The charge current operator along direction $x$ at a given position $\hat{j}_y$ along $y$ is given by
\be
\hat{j}_c(j_y) = \frac{2e}{\hbar} \sum_{k_x} 
{\boldsymbol \psi}_{k_x,j_y}^{\dagger} 
\left(\begin{array}{cc}
-\tilde{t} \sin(k_x) & -\frac{i}{2} \alpha \cos (k_x) \\
\frac{i}{2} \alpha \cos(k_x) & -\tilde{t} \sin (k_x) 
\end{array} \right)
{\boldsymbol \psi}_{k_x,j_y} 
\ee
where ${\boldsymbol \psi}_{k_x,j_y}^{\dagger} =
\left( \psi_{k_x,j_y,\ua}^{\dagger}, \psi_{k_x,j_y,\da}^\dagger \right)$.
The current has contributions from the hopping and the spin-orbit terms.

One may also define a longitudinal spin current, $\hat{j}_s(j_y)$, taking the difference between
the two diagonal components of the charge current. The other terms correspond to
spin-flip terms and do not contribute to the $z$ component of the spin current. 

The average value of the charge current in the groundstate is given by summing over the
single particle occupied states (negative energies) in the usual way
\bea
j_c(j_y) = \langle \hat{j}_c(j_y) \rangle = \sum_{k_x,n} \nonumber \\ 
 \left\{ 
\tilde{t} \sin k_x \left[ 
  \tilde{v}_n(-k_x,j_y,\uparrow) \tilde{v}_n^*(-k_x.j_y,\uparrow) \right. \right.  \nonumber \\
+ \left. \tilde{v}_n(-k_x,j_y,\downarrow) \tilde{v}_n^*(-k_x.j_y,\downarrow) \right] \nonumber \\
- \frac{i \alpha}{2} \cos k_x \left[ 
 \tilde{v}_n(-k_x,j_y,\uparrow) \tilde{v}_n^*(-k_x.j_y,\downarrow) \right. \nonumber \\
- \left. \left. \tilde{v}_n(-k_x,j_y,\downarrow) \tilde{v}_n^*(-k_x.j_y,\uparrow)  \right]
\right\}
\eea
Here the functions are of the type
\be
\tilde{u}_n(k_x,j_y,\sigma) = \sum_m e^{imwt} u_{n,m}(k_x,j_y,\sigma)
\ee
where as usual $\sigma=\uparrow,\downarrow$.

It is also interesting to consider the magnetization of each momentum value. The spin polarization
is obtainable by
\be
\hat{m}(k_x,j_y) = \psi_{k_x,j_y,\uparrow}^{\dagger} \psi_{k_x,j_y,\uparrow} -
\psi_{k_x,j_y,\downarrow}^{\dagger} \psi_{k_x,j_y,\downarrow} 
\ee
and the average spin polarization in the groundstate is given by
\bea
m(k_x,j_y) = \sum_n\nonumber \\ 
\left\{ 
- \tilde{v}_n(-k_x,j_y,\uparrow) \tilde{v}_n^*(-k_x.j_y,\uparrow) \right. \nonumber \\
+ \left. \tilde{v}_n(-k_x,j_y,\downarrow) \tilde{v}_n^*(-k_x.j_y,\downarrow)  \right\} 
\nonumber \\
\eea

\subsection{Charge currents}

\begin{figure}
\includegraphics[width=0.9\columnwidth]{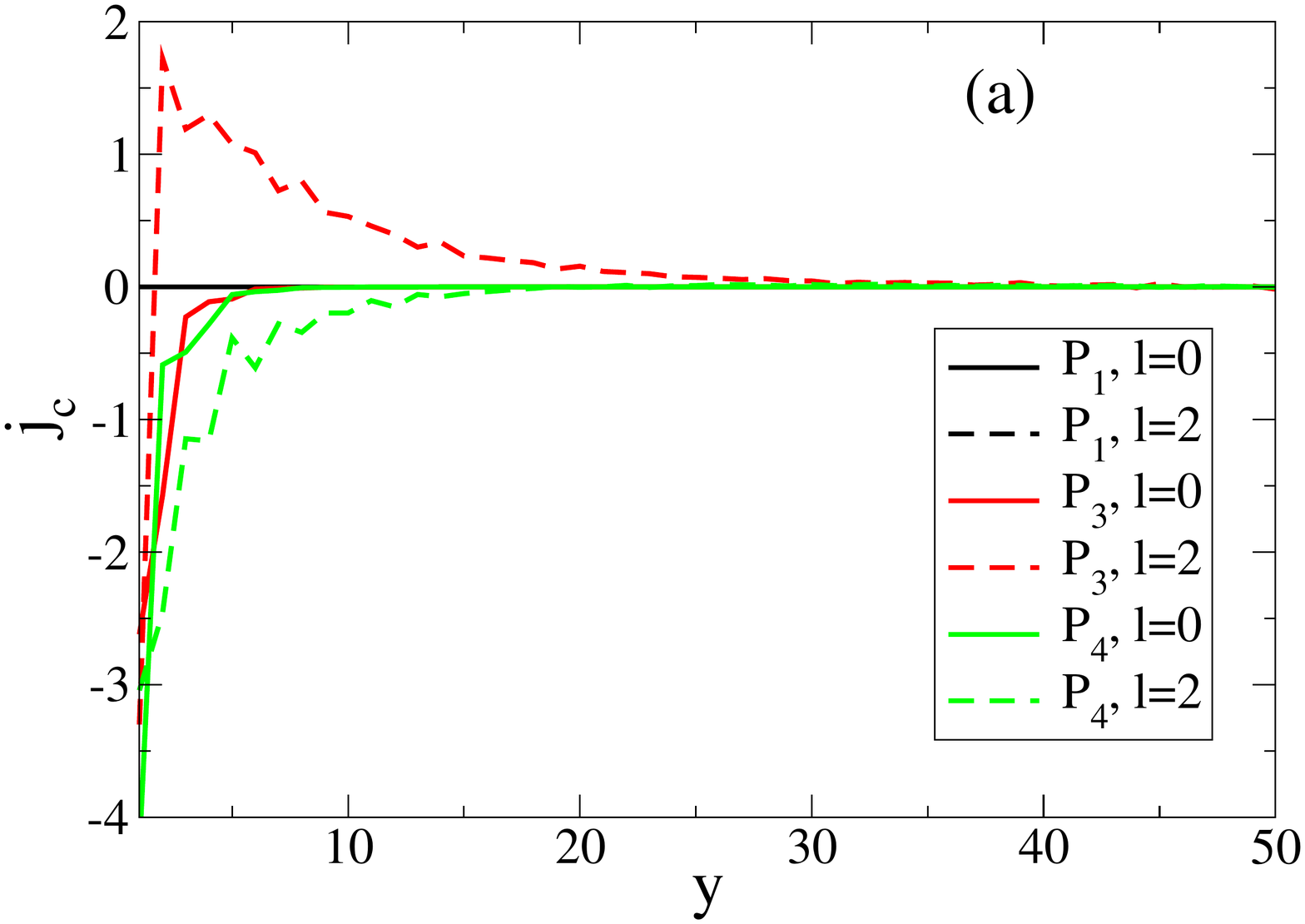}
\includegraphics[width=0.9\columnwidth]{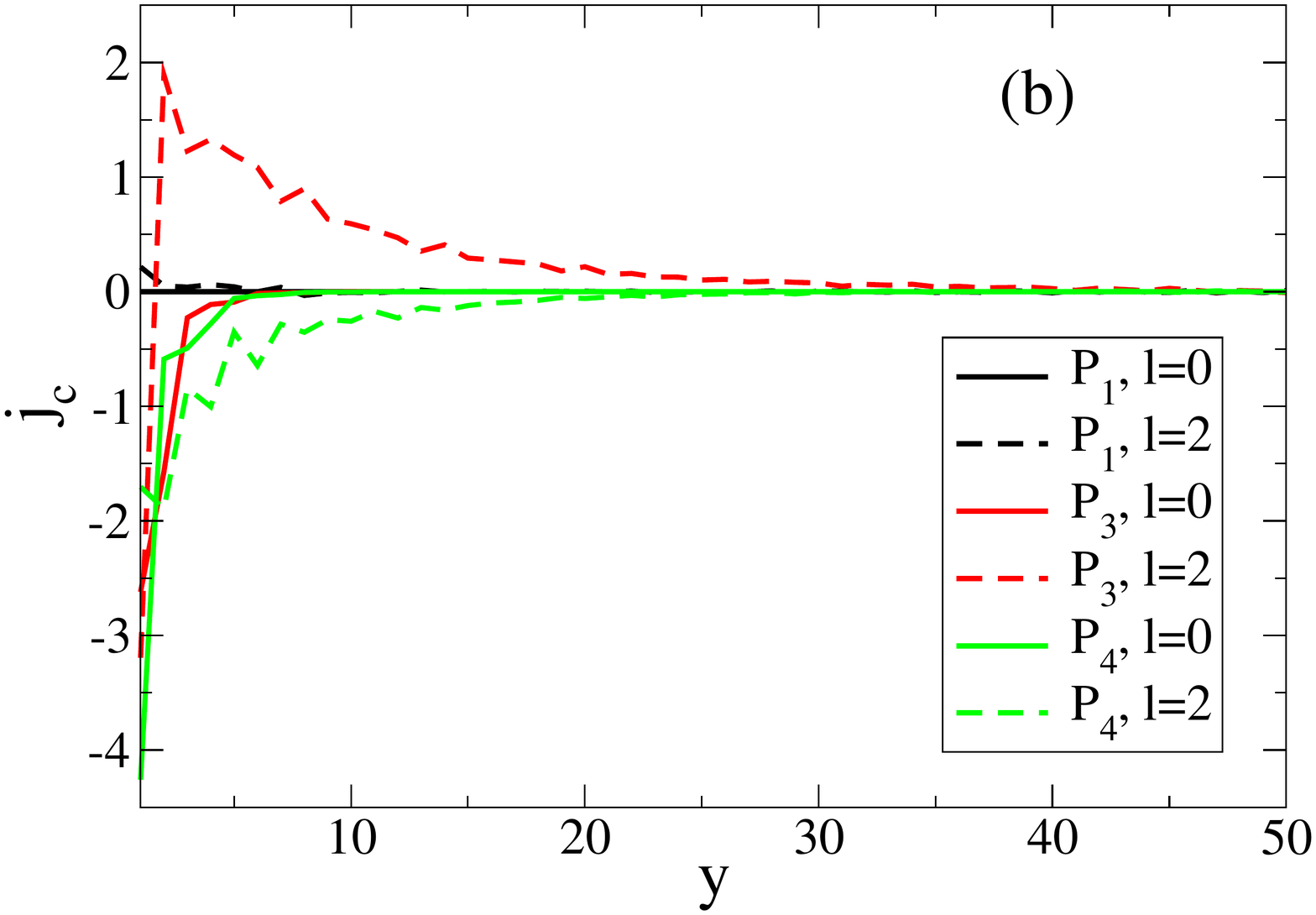}
\includegraphics[width=0.9\columnwidth]{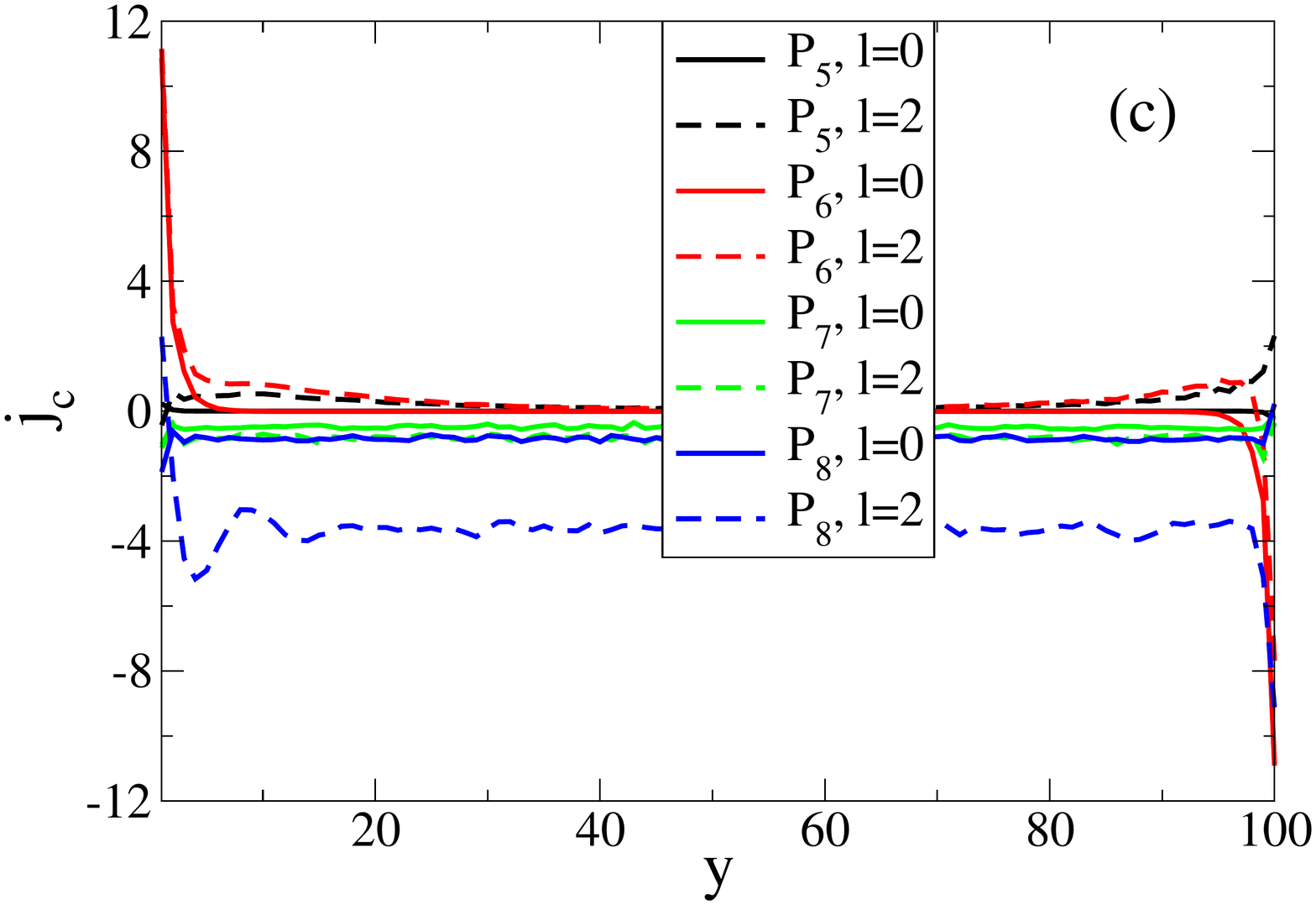}
\caption{\label{fig6}
(Color online) Charge current profile for $l=2$ for a $d_x,d_y$ superconductor
for $P_1,P_3,P_4$ for (a) $\mu_d=1$ and (b) $M_{zd}=1$.
Only one half of the system is shown since the current profile is anti-symmetric
around the middle point. The system size is $N_y=100$.
Charge current profile for $l=2$ for a $d_z$ superconductor
for $P_5,P_6,P_7,P_8$ for (c) $\mu_d=1$.
Note that the current profile has no symmetry around the middle point in
some cases. 
The results for the $d_x,d_y$ pairing consider three points in the phase diagram:
$\mu=-5,M_z=0$ (case $P_1$ with $C=0$), 
$P_3$ with $\mu=-5,M_z=2$ (with $C=1$) and $P_4$ with $\mu=-1,M_z=2$ (with $C=-2$).
The results for the $p+ip$ pairing consider $P_5$ with $\mu=-5,M_z=0$ (with $C=0$),
$P_6$ with $\mu-2,M_z=0$ (with $C=2$), $P_7$ with $\mu=-3,M_z=2$ (with $C=1$) and
$P_8$ with $\mu=-1,M_z=2$ (with $C=-2$). These points are indicated in Fig. \ref{fig1}.
} 
\end{figure}

The charge currents at the edges of the unperturbed Hamiltonian are well
understood. If there is TRS the currents vanish and if TRS is broken the
edge charge currents are finite (finite Chern number). We consider as examples
a $d_x,d_y$ triplet superconductor and a $p+ip$ triplet superconductor.
In the first case, if the magnetization vanishes, the system has TRS and vanishing
charge edge currents in the topologically trivial phases. The charge current
also vanishes in the $Z_2$ topological
phase but the spin edge currents are non-vanishing.
In the case of the $p+ip$ triplet superconductor there is no TRS and the charge
edge currents are finite.

We consider in the case of the $d_x,d_y$ pairing a set of parameters
$d=0.6,\Delta_s=0.1,d_z=0,\alpha=0.6$ and different values for the chemical
potential and the magnetization. In the case of the $p+ip$ pairing we consider
$d=0,\Delta_s=0.1,d_z=0.6,\alpha=0.1$.

In Fig. \ref{fig6} we show results for the profile of the charge current as a function of $y$
for the various cases. We compare the unperturbed case with the perturbed one
by considering that at $w=6$ it is enough to truncate the
Hamiltonian matrix at $l=2$. To calculate the currents we sum over the states
in the first Floquet zone. Also the results are for time $t=0$ or any multiple
of the time period $T$.

The results for the $d_x,d_y$ pairing consider four points in the phase diagram: namely
a trivial phase $\mu=-5,M_z=0$ (case $P_1$ with $C=0$), a $Z_2$ phase $\mu=-3,M_z=0$
(case $P_2$ with $C=0$), 
and two points in $Z$ topological phases
$P_3$ with $\mu=-5,M_z=2$ (with $C=1$) and $P_4$ with $\mu=-1,M_z=2$ (with $C=-2$).
The results for the $p+ip$ pairing consider $P_5$ with $\mu=-5,M_z=0$ (with $C=0$),
$P_6$ with $\mu-2,M_z=0$ (with $C=2$), $P_7$ with $\mu=-3,M_z=2$ (with $C=1$) and
$P_8$ with $\mu=-1,M_z=2$ (with $C=-2$). These points are indicated in Fig. \ref{fig1}.

In  Fig. \ref{fig6}a we consider a periodic driving in the chemical potential
with amplitude $\mu_d=1$ and in  Fig. \ref{fig6}b in the magnetization with the same amplitude.
Focusing first on the unperturbed case we see that the current at the border vanishes for
$P_1$ ($C=0$) and that for $P_4$ with $C=-2$ the current is larger at the edge, but decays faster with distance from the
border with respect to the $P_3$, $C=1$ case. Adding the perturbations in the case of the trivial $P_1$ case,
there is no charge current if driving the chemical potential but if one drives the magnetization
there is a small current close to the edge. The effect of the perturbations on the topological
cases is however quite significative, particularly in the case of $C=1$, with a current inversion with
respect to the unperturbed case. 
A similar result has been found in the context of $d_{x^2-y^2}+i d_{xy}$ pairing \cite{did}.
In both cases the spatial extent of the currents away from the border
is considerably increased. However, we have checked that integrating the currents over $y$ from the edge
to the middle point the total current is approximately the same. Also, we see that if the unperturbed
state is topologically non-trivial the effect of the two types of perturbations is very similar.

In  Fig. \ref{fig6}c we show results for the $p+ip$ pairing. Except for the case
with $C=0$, the charge currents are finite in the unperturbed and perturbed cases. We note that,
even for the $C=0$ case, there is a small charge current at the border, probably due to some finite
size effects. In general, the driven chemical potential considered here increases the charge current.
In the cases with zero magnetization the current decreases from the borders but at finite $M_z$ the
current is finite throughout the whole system and is not symmetric around the middle point.

\subsection{Spin currents}

\begin{figure}
\includegraphics[width=0.9\columnwidth]{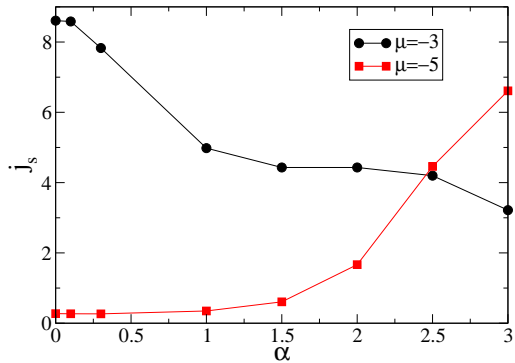}
\caption{\label{fig7}
(Color online) Influence of the spin-orbit coupling on the spin current in the unperturbed
case. The parameters are $d=0.6,\Delta_s=0.1, d_z=0, M_z=0, \mu=-3,-5$. Increasing
the spin-orbit in the topologically non-trivial phase is detrimental while in the
topologically trivial phase, increasing the spin-orbit coupling considerably increases
the spin current.} 
\end{figure}

\begin{figure}
\includegraphics[width=0.9\columnwidth]{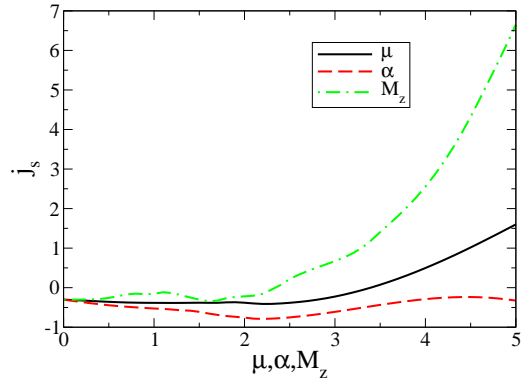}
\caption{\label{fig8}
Influence of intensity of periodic driving for the cases of $\mu_d,\alpha_d,M_{zd}$ on
the spin current ($l=2$). The parameters are $\mu=-5,d=0.6,\Delta_s=0.1,d_z=0,M_z=0,\alpha=0.6$
and the frequency of the periodic driving is $w=6$. Note the reversal of the direction of the
spin current at small couplings. At weak coupling the increase of the
spin current is linear. As the transition from weak to strong coupling occurs the spin current
becomes non-linear. Changing the spin-orbit coupling has a small effect on the edge spin
current while both changing the chemical potential and the magnetization have strong
effects includind a change of the current direction.
} 
\end{figure}

\begin{figure*}
\includegraphics[width=0.45\textwidth]{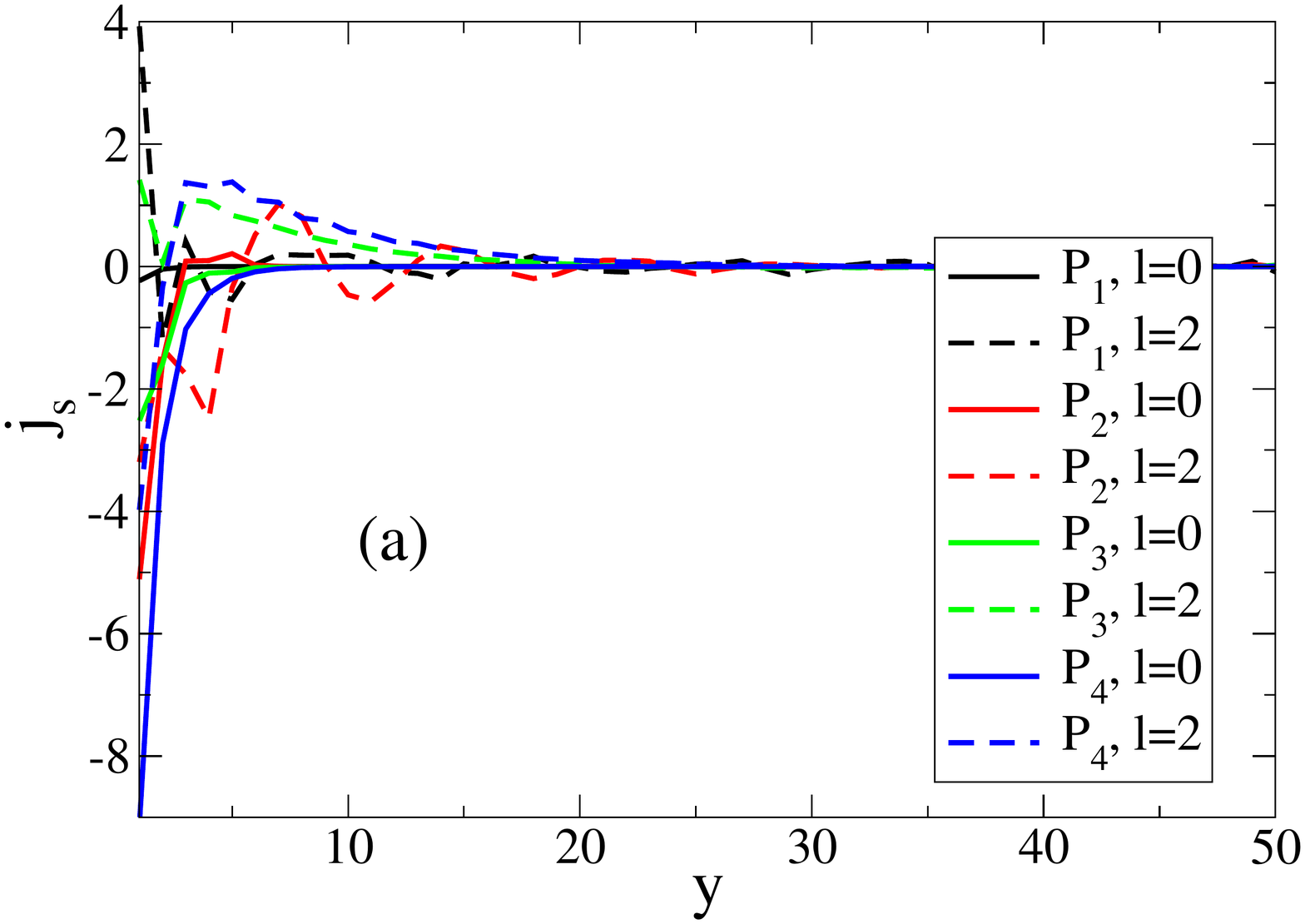}
\includegraphics[width=0.45\textwidth]{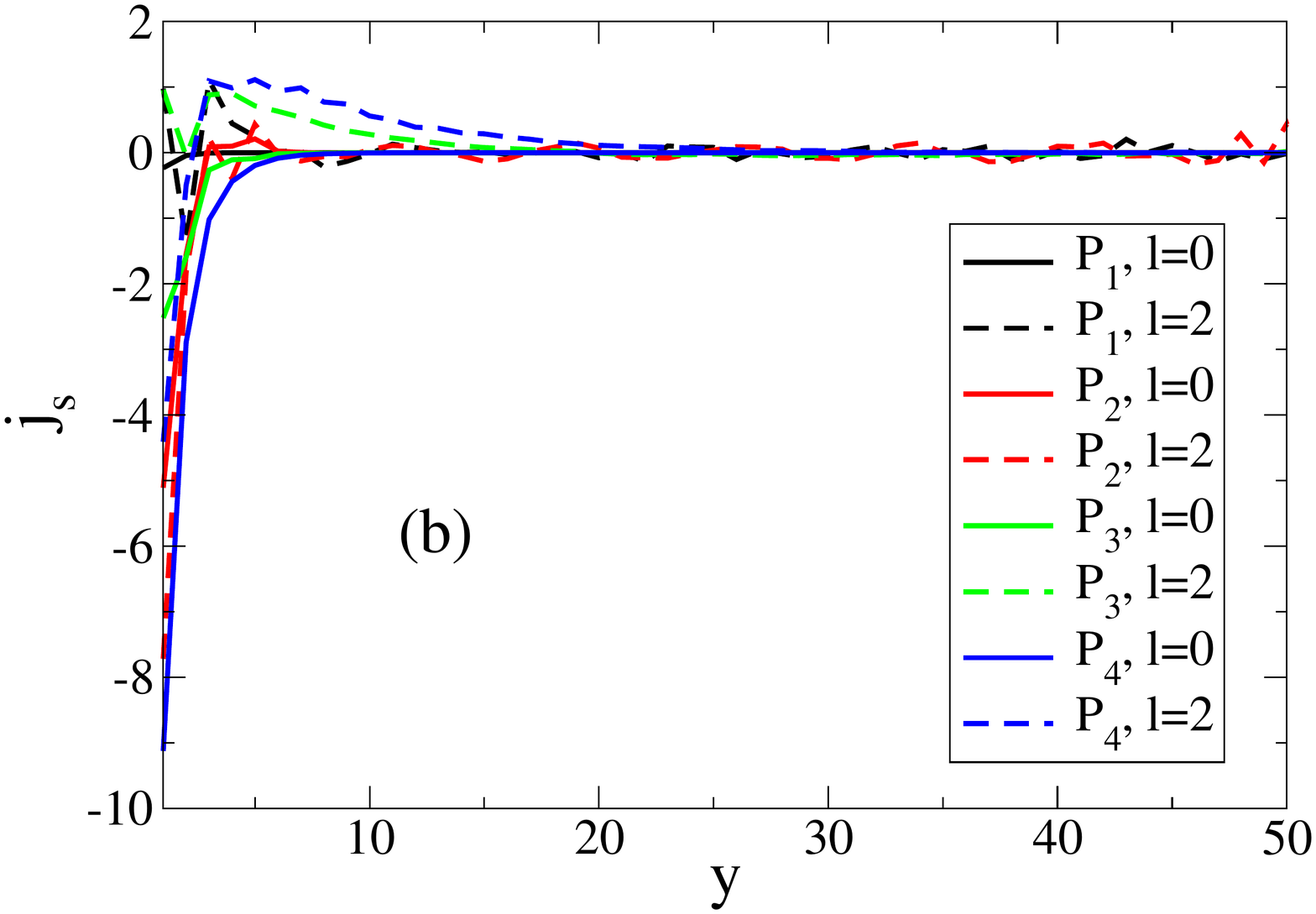}
\includegraphics[width=0.45\textwidth]{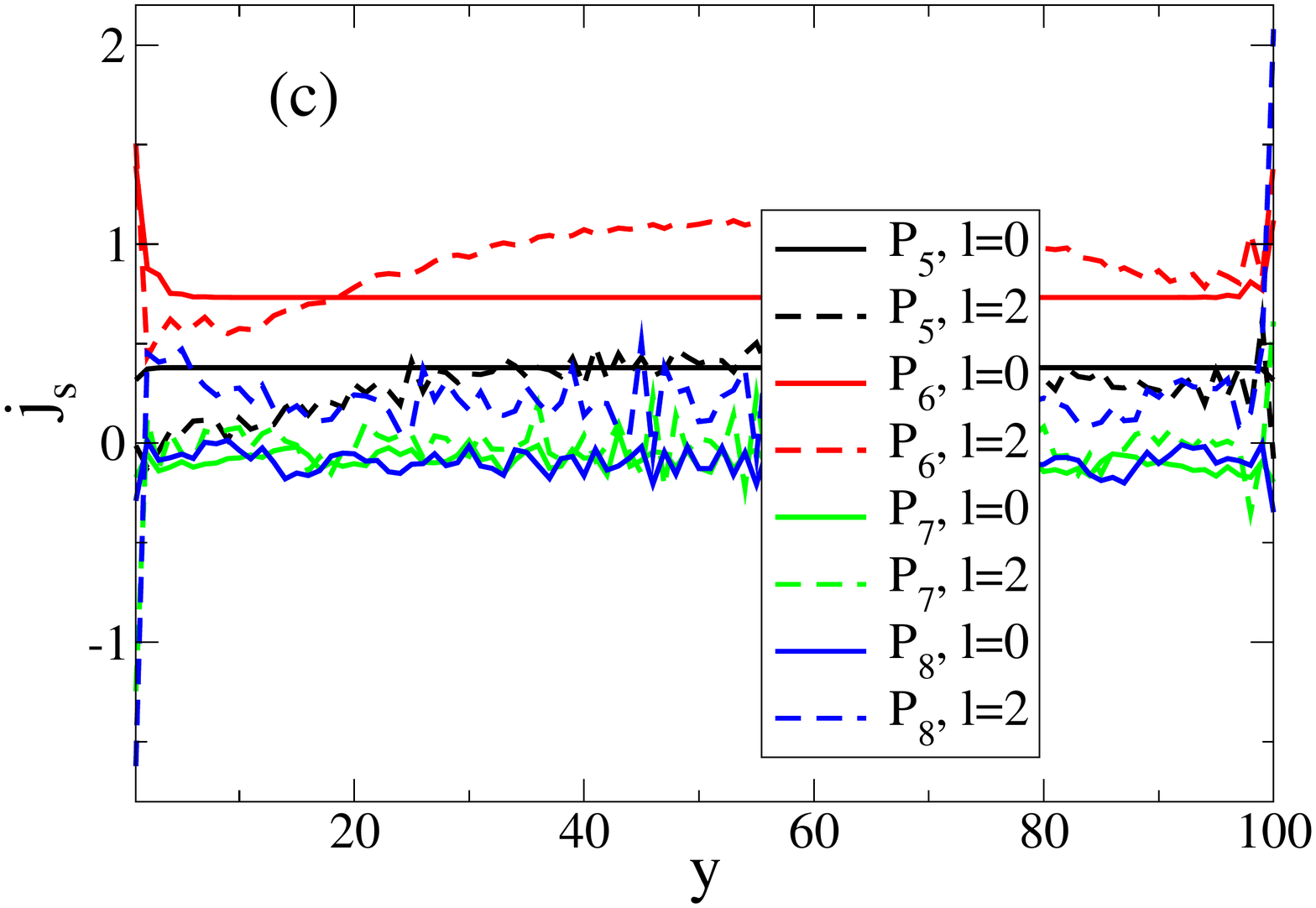}
\includegraphics[width=0.45\textwidth]{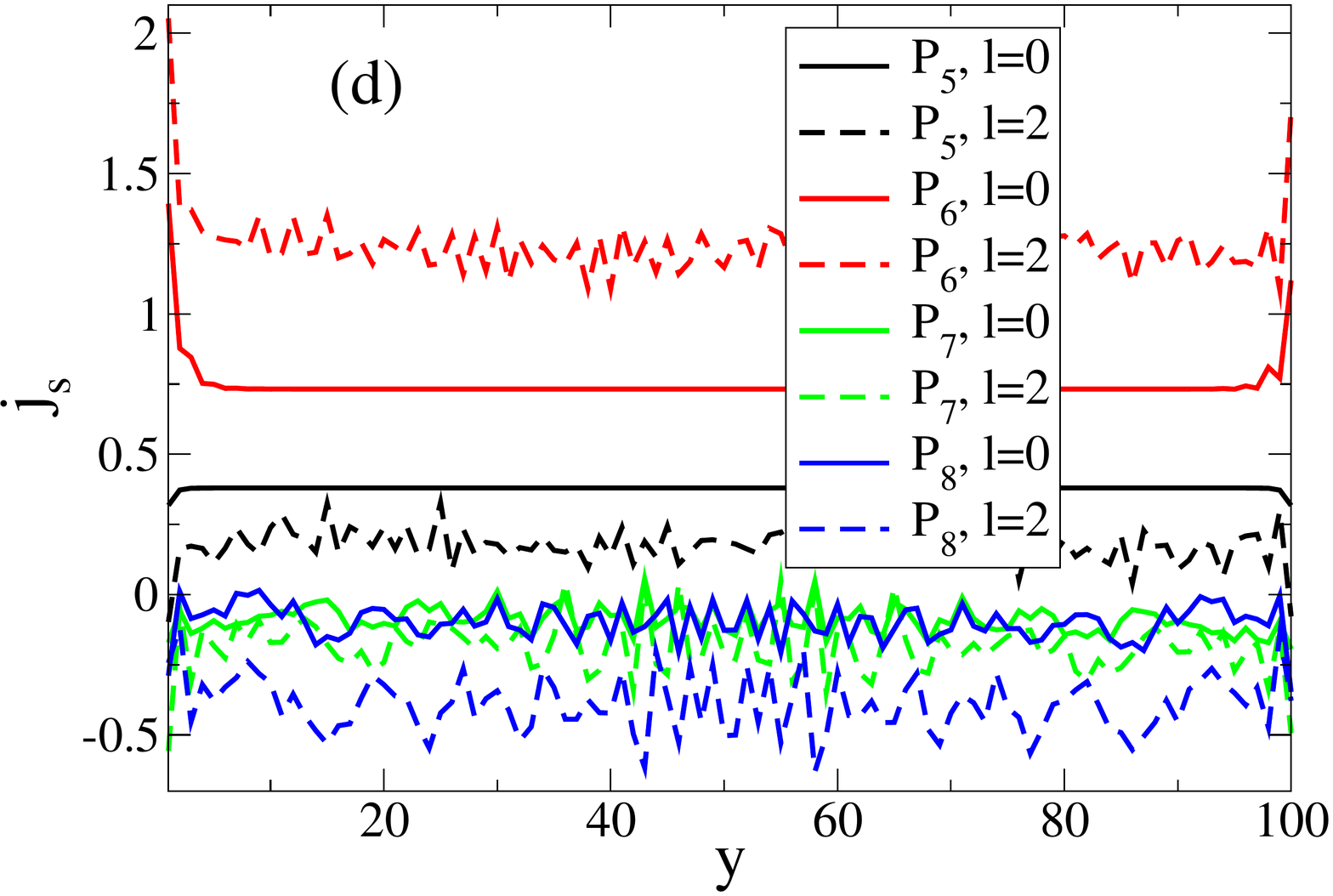}
\caption{\label{fig9}
(Color online)
Spin current for the triplet superconductors $d_x,d_y$ ((a), (b)) and
$d_z$ ((c), (d)) for the periodic drivings $\mu_d=1$  ((a), (c)) and
$M_{zd}=1$ ((b), (d)).
} 
\end{figure*}

In this subsection we calculate the longitudinal spin currents at the edges of the system,
both for the unperturbed and perturbed systems.

Consider first the unperturbed triplet superconductor.
We start with the $d_x,d_y$ triplet superconductor in a trivial phase in
zero magnetic field. In this regime there are no edge states and the system has
a finite gap around zero energy. In Fig. \ref{fig7} we consider the influence of
the spin-orbit coupling on the spin current in this trivial phase and compare it
with the case of a topological $Z_2$ phase, where there are edge states that, due
to the spin locking, counterpropagate at each edge yielding a null charge current but a finite
spin current. This is shown in the results for the case with $\mu=-3,M_z=0$.
Increasing the spin-orbit coupling decreases the longitudinal spin current.
An opposite effect is observed in the trivial phase for which, as the spin-orbit coupling
increases, the spin current increases. At very small coupling the spin current
is very small but non-vanishing, probably due to finite size effects. Even though there are
no edge states the spin current is also carried out by the states in the continuum
above and below the gap. When there are edge states these carry most of the current,
but there is also a contribution from the other states in the continuum.

\begin{figure*}
\includegraphics[width=0.32\textwidth]{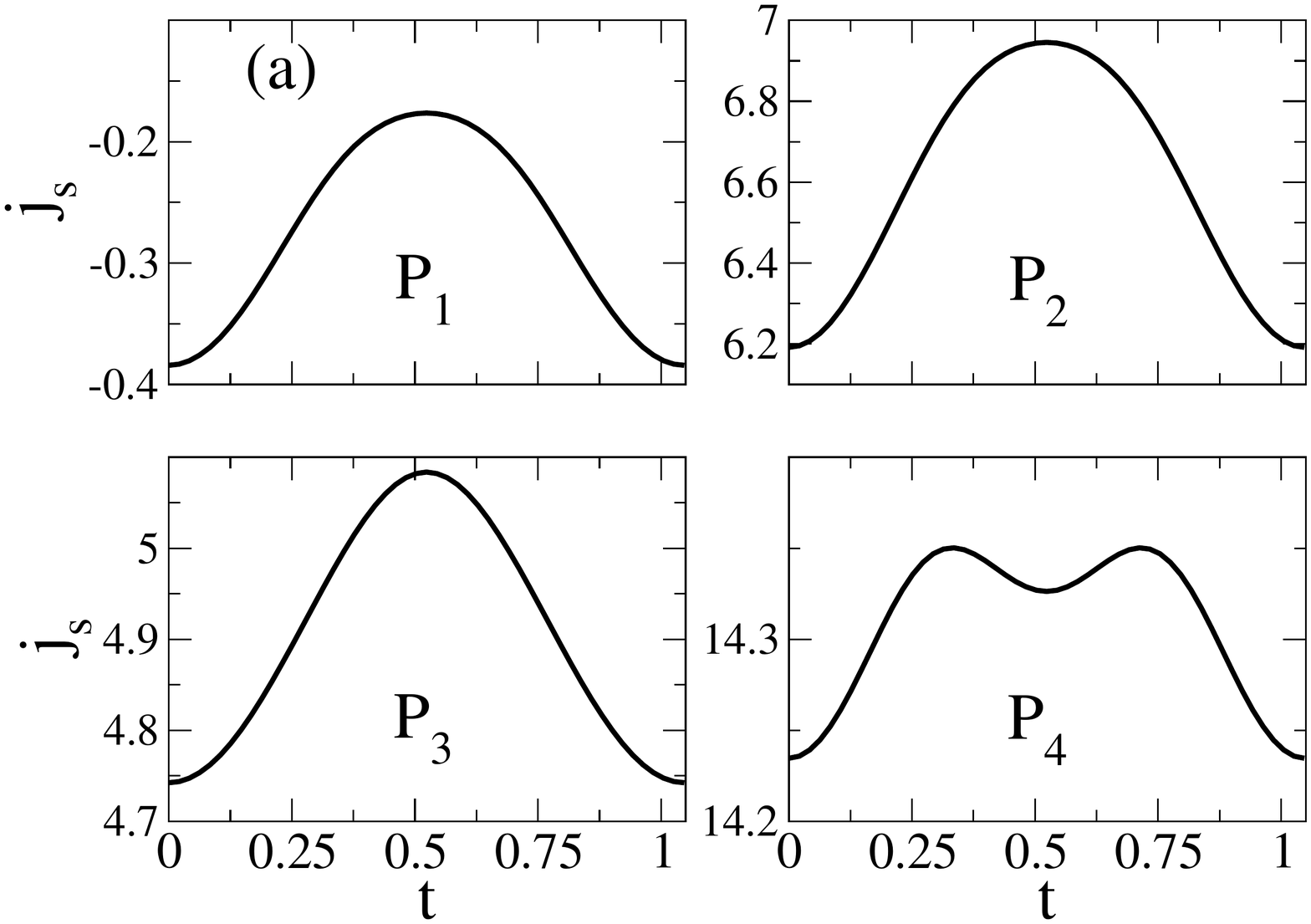}
\includegraphics[width=0.32\textwidth]{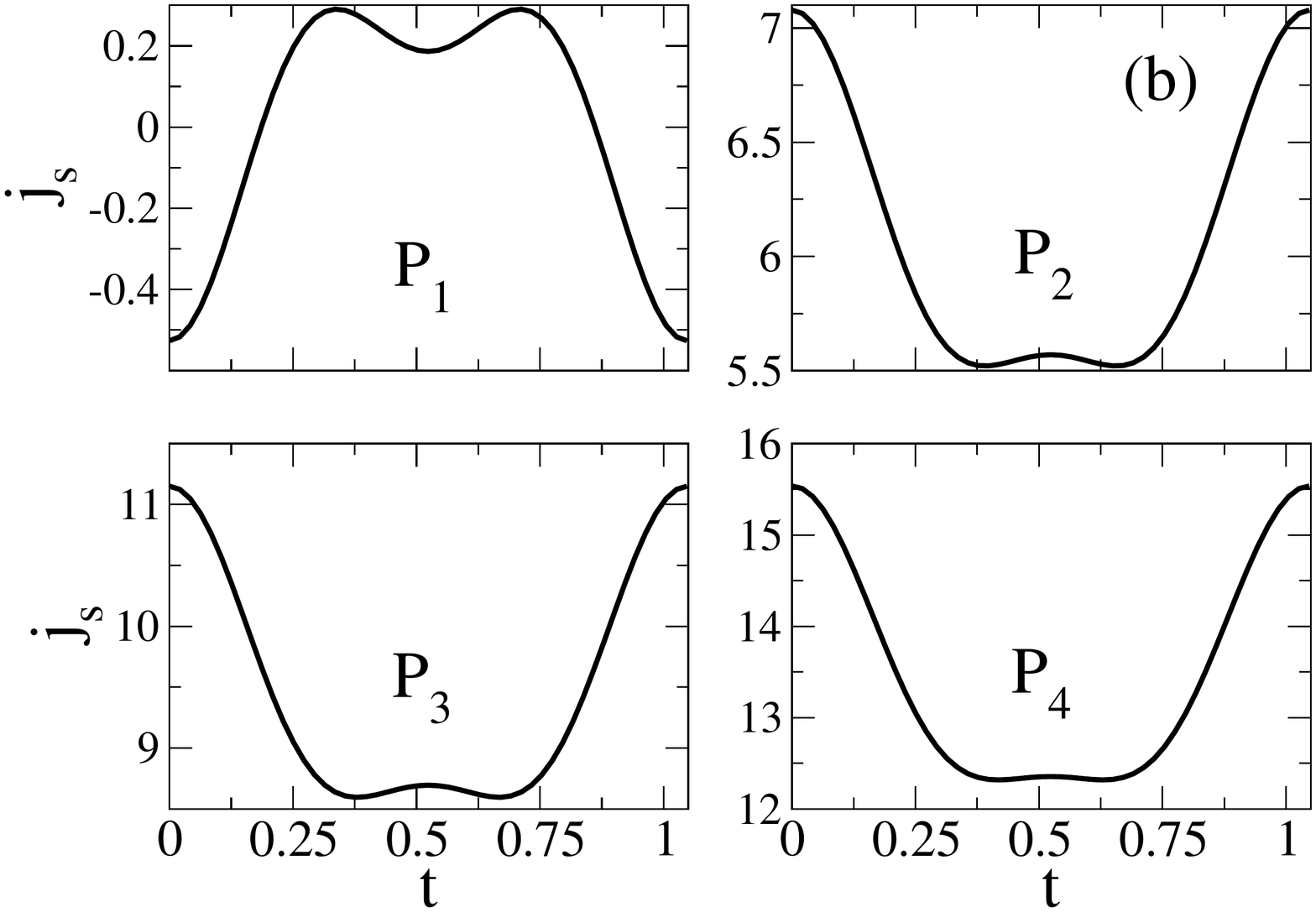}
\includegraphics[width=0.32\textwidth]{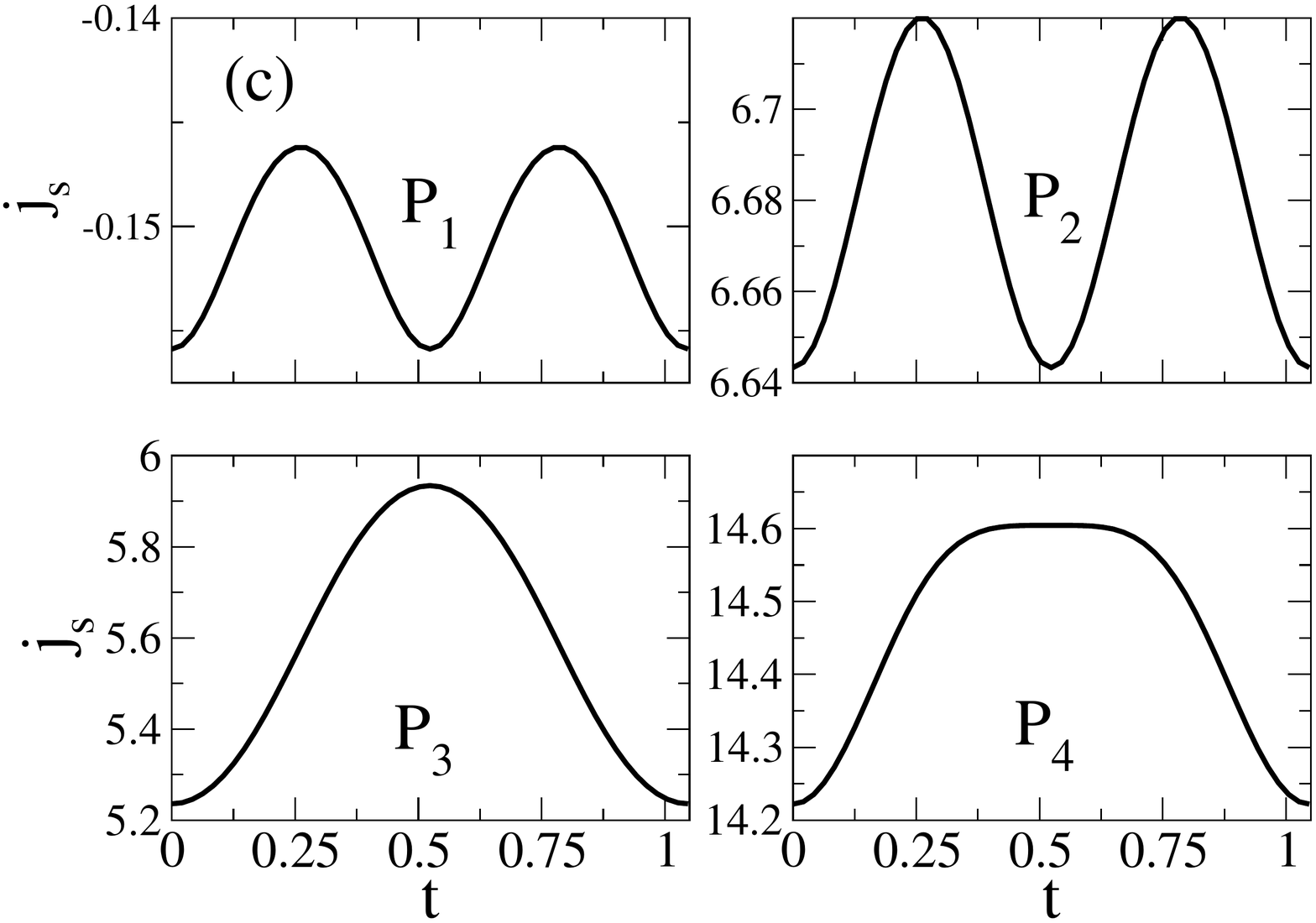}
\caption{\label{fig10}
Time evolution of the spin current for a $d_x,d_y$ triplet
superconductor for the periodic drivings (a) $\mu_d$, (b) $\alpha_d$, and (c) $M_{zd}=1$
for $P_1,P_2,P_3,P_4$, respectively.
} 
\end{figure*}

Consider now the perturbed cases.
There are two parameters that characterize the perturbation: the frequency of the periodic
driving and its amplitude. 
As discussed above, small frequencies require a large matrix to be diagonalized, particularly
to determine the edge states. Very large frequencies are amenable to analytical solutions
using the Magnus expansion and can be interpreted as a renormalization of the parameters
of the original Hamiltonian, or by the addition of new terms allowed in the original
unperturbed Hamiltonian. The cases treated here respect to intermediate values of the
frequency, to allow for a simple truncation scheme and, at the same time, a resonance condition
between bands
that is most interesting in changing the topology of the system.
Most results reported here are for a small to moderate amplitude.
If the perturbation has a very small amplitude, it can be dealt with using perturbation theory.
Very large amplitudes deviate considerably the system from its initial state. Therefore we have
considered here moderate amplitudes. To illustrate the effect of the increase of the perturbation
amplitude we show in Fig. \ref{fig8}, for the same triplet superconductor, the effect
of changing the amplitude of a perturbation in the chemical potential, spin-orbit coupling and
magnetization. As above for the charge currents, the results hold for time $t=0$ or for any
multiple of the period $T$. 
The results for the
spin current are obtained integrating the local spin current accross half the sample. 

In all cases there is a sign reversal of the spin current.
At small couplings, the behavior of the spin current ($l=2$) is approximately
linear and the spin current does not change appreciably. Increasing the amplitude in the case
of the spin orbit coupling, $\alpha_d$, does not affect much the spin current even though the
response is no longer linear. On the other hand, the change in the chemical potential leads
to a significant change of the spin current, changing back its direction for large enough values of the
chemical potential, $\mu_d$, and the effect is even more pronounced in the case of the magnetization, $M_{zd}$.
These results were obtained taking the unperturbed as a trivial phase with a very small spin current.
The spin current is therefore significantly enhanced by the periodic driving. 

In Fig. \ref{fig9} we show results for $t=0$ (or a multiple of the period) of the spatial
profile of the spin current for the triplet superconductors $d_x,d_y$ ((a) and (b)) and
$p+ip$ ((c) and (d)) obtained changing the chemical potential, $\mu_d$, or the magnetization,
$M_{zd}$.
In the case of the first pairing symmetry, the spin current is antisymmetric around the middle point
and so it is enough to consider half of the system. In the case of the $p+ip$ superconductor
the current has no symmetry and so the full profile is shown, as before for the charge current.

The results are to some extent qualitatively the same as for the charge current, in the sense
that the periodic driving induces a large spin current, if the unperturbed system is
trivial, and enhances the spin current if the system is originally topological.
The spin current oscillates, there is also reversal of direction due to the periodic
driving, in the case of $C=-2$, and the spatial extent of the current profile increases.
As for the unperturbed case, when there are edge states, the largest contribution to the
current originates in them, but the continuum states also contribute.

In the case of the $p+ip$ superconductor the currents are spread along the whole system and
in general are not symmetric around the middle point of the $y$ direction.

At arbitrary time values the currents oscillate. In Fig. \ref{fig10} we show the
time evolution of the spin current within one period, $T$, for the case of the
$d_x,d_y$ triplet superconductor driven by the chemical potential, spin-orbit and
magnetization. We consider as unperturbed states in each panel the trivial case
$P_1$, the $Z_2$ topological point $P_2$, and the $Z$ topological phases 
$P_3$ and $P_4$.  In the trivial phase, at short times, the perturbations decrease the spin current
with respect to the initial time value (or any multiple of the time period).
In the non-trivial phases the chemical potential and magnetization perturbations increase the spin currents at short times with respect to
the initial time value while the spin-orbit coupling acts the opposite way.

\begin{figure}
\includegraphics[width=0.42\columnwidth]{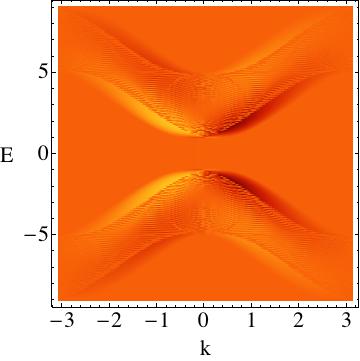}
(a)
\includegraphics[width=0.42\columnwidth]{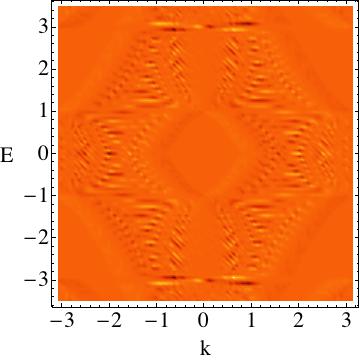}
(b)
\includegraphics[width=0.42\columnwidth]{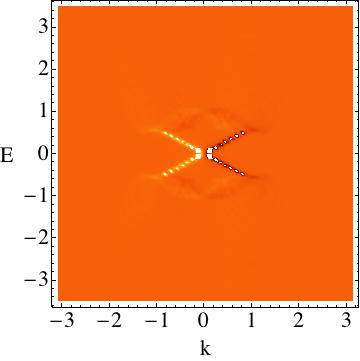}
(c)
\includegraphics[width=0.42\columnwidth]{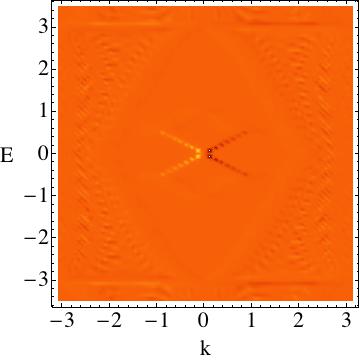}
(d)
\caption{\label{fig11}
(Color online) Spin polarization of unperturbed ($l=0$) and perturbed ($l=2$) system with $\mu_d=1$ 
for (a) and (b) $P_1$ (trivial case)  and (c) and (d) $P_2$ ($Z_2$ case). In these cases $C=0$ and $M_z=0$.  
The maximal values of the spin polarizations are
(a) $7.3\times 10^{-5}$, (b) $1.9\times 10^{-3}$, (c) $2.5\times 10^{-2}$ and (d) $5.8\times 10^{-3}$.
The maximal value displayed in panel (c) has been reduced to $0.007$ for better visualization.} 
\end{figure}

\subsection{Spin polarization}

Associated with the spin currents in the topological phases, it has been
shown that the eigenstates have non trivial spin polarizations that 
depend strongly on the momentum \cite{timm}.  It was shown that, particularly
for cases where one has flatbands, there is a strong polarization effect. 
Here we will look at the spin polarization
of the Floquet states and compare with the unperturbed cases.

\begin{figure}
\includegraphics[width=0.42\columnwidth]{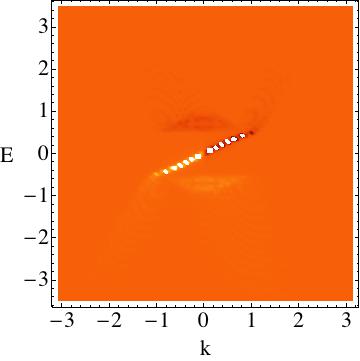}
(a)
\includegraphics[width=0.42\columnwidth]{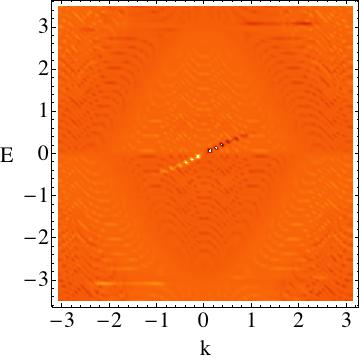}
(b)
\includegraphics[width=0.42\columnwidth]{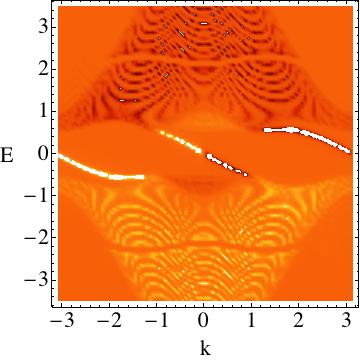}
(c)
\includegraphics[width=0.42\columnwidth]{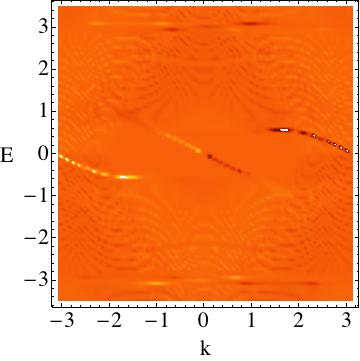}
(d)\caption{\label{fig12}
(Color online) Spin polarization of unperturbed ($l=0$) and perturbed ($l=2$) system with $\mu_d=1$ for
(a) and (b) $P_3$ ($C=1$) and (c) and (d) $P_4$ ($C=-2$). In these cases $C\neq 0$ and $M_z \neq 0$.
The maximal values of the spin polarizations are
(a) $3.8\times 10^{-2}$, (b) $7.8\times 10^{-3}$, (c) $4.7\times 10^{-2}$ and (d) $8.3\times 10^{-3}$.
The maximal values of the panels have been reduced to $0.005$.
} 
\end{figure}

In Figs. \ref{fig11} and \ref{fig12}
we show the spin polarization as a function of momentum for each quasi-energy state
for the various examples, $P_1,P_2,P_3,P_4$ of the $d_x,d_y$ superconductor, both for the unperturbed and the perturbed cases
with $l=2$.

The unperturbed cases in Fig. \ref{fig11} show the nontrivial momentum distribution
of the spin polarization for cases where $C=0$ and $M_z=0$. 
Summing over all momenta and states, the total polarization
vanishes. The spin polarization shown in Fig. \ref{fig11}a has higher values at the 
edges of the gap and opposite signs as the momentum changes sign, due to the
time reversal symmetry. 
Note, however, that the spin polarization is very small ($\sim 10^{-5}$).
In this regime of parameters the system is in a trivial phase.

Adding the periodic driving edge states are generated,
as shown above, the spin polarization has higher values that are more uniformly spread over the various
energy states and momenta, and with clearly higher values at the induced edge states
at the border of the Floquet zone, again with opposite signs as the momentum
changes sign since time reversal symmetry is not broken. 
However, the spin polarizations have a complex structure over the continuum.
Panels (c) and (d) refer to a phase that shows edge states already at the level of the unperturbed
Hamiltonian but with time reversal symmetry. 
The spin polarization is much stronger and sharply localized along the states that constitute the edge modes.
Turning on the perturbation, the weight along the original edge states remains evident, but is reduced and there is
a pile up of spin polarization along the new Majoranas generated at the edge of the Floquet zone.
Note that, even though there is an alternancy of the polarization sign when the momentum sign is reversed,
the two edge states are quite visible.

The results in Fig. \ref{fig12} are different. They apply to cases for which the Chern number does not vanish,
TRS is broken in the unperturbed system and there is a finite magnetization. 
The spin polarization is once again concentrated mostly along
the edges but only one edge state is visible. The edge may be switched to the other one by reversing the sign of the
magnetization. 
If the edge states are present in the unperturbed case, it is also clear that adding the time
perturbaton decreases the polarization overall amplitude.
Considering unperturbed topological phases, the spin polarization is highly concentrated
on the edge states and less so in the perturbed case. In any case there is an enhancement
with respect to the bulk states along the edge states, with either positive or negative spin polarizations in both branches
of the edge states. However, in the presence of a magnetic field this symmetry is gone.

In the case of the $p+ip$ superconductor (not shown here)
the states of the unperturbed Hamiltonian below the gap have a
spin polarization that is large (and positive) for the negative momenta values but is
considerably smaller for the positive momenta. The opposite happens for the
positive energy states. 
Both cases correspond to zero magnetization but the chiral nature is evident, both in
the trivial case $C=0$ (expressed in the bulk states) and in the topologically non-trivial
phase ($C=-2$), (also in the edge states). 
This effect is somewhat washed out turning on the periodic driving, but
both for the unperturbed and
perturbed cases, only one edge state has a large spin polarization.
In the perturbed case the edge state is immersed in the continuum. Nevertheless
the spin polarization clearly singles it out.

\section{Conclusions}

A periodic driving on a topologically trivial system induces edge modes
and topological properties. In this work we considered triplet and singlet superconductors subject
to periodic variations of the chemical potential, spin-orbit coupling and magnetization
in both topologically trivial and nontrivial phases and studied their influence on the
charge and spin currents that propagate along the edges of the two-dimensional system.
The most relevant case is the generation of the charge and spin edge currents due to the periodic
driving in an otherwise trivial system. 
In some cases the generated currents are quite high and higher than in the unperturbed case
providing a way to generate and control high spin currents with potential interest in the
context of spintronics, particularly since some of the driving protocols involve electric
or gate potential means, instead of a direct magnetic manipulation.
Starting from a topologically non-trivial phase,
the edge states originally present at zero energy are complemented by edge states at the edge
of the Floquet zone at finite quasi-energies close to $\pm w/2$. In general, the periodic driving
smears the edge states, as shown particularly in the study of the spin polarization of the
quasi-energy states, with a complex structure both in the unperturbed and perturbed cases.
The chirality of the edge states is particularly seen when TRS is broken either by the
presence of a finite magnetization or due to the intrinsic TRS breaking in a $p+ip$ triplet
superconductor.

The transport signatures of edge states in driven topological insulators or due to the 
Floquet Majorana fermions in driven topological superconductors,
involve the presence of leads to inject and collect charge or spin currents and therefore
require the coupling of the system to outside reservoirs, with the associated issue of
dissipation \cite{mitra2}. A calculation of the differential conductance has found a quantized 
conductance sum rule \cite{conductance}, which
generalizes the quantized zero-bias conductance contribution of the Majorana fermions
when the system is not periodically driven.  
Here we neglected any couplings to external reservoirs.
The charge currents calculated in this work may be detected considering the magnetic field
they generate in their vicinity, as proposed in the context of unperturbed systems in Ref. \cite{moore3}.  
Barring in mind the difficulty in detecting experimentally the edge currents, it may be challenging
as well to detect the generated currents considered here. 

Discussions with Xiaosen Yang, Maxim Dzero, Antonio Garcia-Garcia and Pedro Ribeiro
are acknowledged. Partial support in the form of a BEV by the CNPq at CBPF is gratefully
acknowledged as well as support from FCT through grant UID/CTM/04540/2013.

\end{document}